\documentclass{PoS}
\def\be{\begin{equation}}
\def\ee{\end{equation}}
\def\bea{\begin{eqnarray}}
\def\eea{\end{eqnarray}}
\def\ba{\begin{array}}
\def\ea{\end{array}}

\newcommand{\xbf}[1]{\mbox{\boldmath $ #1 $}}

\title{Quadrupole moment of the nucleon in chiral constituent quark model}

\ShortTitle{Quadrupole moment of the nucleon}

\author{\speaker{Harleen Dahiya}
\thanks{H.D. would like to thank
the organizers of Light Cone 2010 and Department of Science and
Technology, Government of India, for financial support.}\\
        Dr. B.R. Ambedkar National Institute of Technology,
        Jalandhar 144011 INDIA \\
        E-mail: \email{harleen\_dahiya@yahoo.com}}

\author{Neetika Sharma\\
Dr. B.R. Ambedkar National Institute of Technology, Jalandhar
144011 INDIA\\ E-mail: \email{neetikaphy@gmail.com}}

\abstract{The electromagnetic form factors are the most
fundamental quantities to describe the internal structure of the
nucleon and the shape of a spatially extended particle is
determined by its {\it intrinsic} quadrupole moment which is first
order moment of the charge density operator. With some
experimental indications of a deformed nucleon, we have calculated
the {\it intrinsic} quadrupole moment of the octet and decuplet
baryons in the framework of chiral constituent quark model
$\chi$CQM which is quite successful in explaining some of the
important baryon properties in the nonperturbative regime.}

\FullConference{Light Cone 2010: Relativistic Hadronic and
Particle Physics\\
        June 14-18, 2010\\
        Valencia, Spain}

\begin{document}

\section{Introduction}
The electromagnetic form factors are the fundamental quantities of
theoretical and experimental interest to investigate the internal
structure of nucleon. The knowledge of internal structure of
nucleon in terms of quarks and gluons degrees of freedom of
Quantum Chromodynamics (QCD) provide a basis for understanding
more complex, strongly interacting matter.  While QCD is accepted
as the fundamental theory of strong interactions, it cannot be
solved accurately in the nonperturbative regime. A coherent
understanding of the hadron structure in this energy regime is
necessary to describe the strong interactions as they are
sensitive to the pion cloud and provide a test for the QCD
inspired effective field theories based on the chiral symmetry.
Recently, a wide variety of accurately measured data have been
accumulated for the static properties of baryons, for example,
masses, electromagnetic moments, charge radii, and low energy
dynamical properties such as scattering lengths and decay rates
etc.. The charge radii and quadrupole moments are important
observables in hadronic physics as they lie in the nonperturbative
regime of QCD and give valuable information on the internal
structure of hadrons.

The Naive Quark Model (NQM) is unable to explain the magnitude and
sign of deformation measured for the case of nucleon and
$\Delta^+$.  A promising approach offered to solve QCD in this
energy regime are the constituent quark models which can be
constructed so as to include the relevant properties of QCD coming
from the consequences of the spontaneous breaking of chiral
symmetry ($\chi$SB). One of the important nonperturbative
approaches in this energy regime is chiral constituent quark model
($\chi$CQM) \cite{wein}. The $\chi$CQM coupled with the ``quark
sea'' generation through the chiral fluctuation of a constituent
quark into a Goldstone bosons (GBs) \cite{cheng,johan,song},
successfully explains the ``proton spin crisis'' \cite{hd},
hyperon $\beta$ decay parameters \cite{nsweak}, the octet and
decuplet baryons magnetic moments \cite{hdmagnetic}. The extension
to the SU(4) symmetry successfully predicts the contribution of
{\it intrinsic} charm (IC) content in the low lying and charmed
baryon magnetic moments and their radiative decay widths
\cite{hdcharm}. In this context, it become desirable to extend the
model to other low energy properties like charge radii, quadrupole
moment, and higher order moments of the multipole expansion.

The purpose of present communication is to calculate the {\it
intrinsic} quadrupole moment of the octet and decuplet baryons
within the framework of $\chi$CQM using the general
parameterization method. In order to understand the important role
played by the pion cloud and SU(3) symmetry breaking in measuring
the quadrupole moment we would carry out the calculations with and
without symmetry breaking. The results have also been compared
with the NQM predictions and the latest available data. Further,
it would also be interesting to understand in detail the role of
GP model parameters in the determination in of baryon quadrupole
moment.

\section{Intrinsic quadrupole moment of the nucleon}
The mean square charge radii $(r^2_B)$ and quadrupole moments
(${Q}_B$) are the lowest order moments of the charge density
operator $\rho$ in a low-momentum expansion. For example, for any
baryon $|B\rangle$ with charge $e_B$, the terms up to order of
$q^2$ for the charge density are \be \langle {\mathrm B}| \rho(q)
|B \rangle = e_B - \frac{q^2}{6} r^2_B - \frac{q^2}{6} {Q}_B +
..... \,,\ee where $q$ is the photon momentum transferred to the
baryon. The first two terms arise due to the spherically symmetric
monopole part of charge density, while the third term arises due
to the quadrupole part of the charge density. They characterize
the total charge, spatial extension of charge radii, and shape of
the system, respectively.

The shape of a spatially extended particle is determined by its
quadrupole moment \cite{buch01,buch04,buch07}, corresponding to
the charge quadrupole form factor $G_{C2}(q^2)$ at zero momentum
transfer. The {\it intrinsic} quadrupole moment of a nucleus with
respect to the body frame of axis is defined as \be Q_0=\int
d^3r~\rho({\bf r}) (3 z^2 - r^2)\,. \ee If the charge density is
concentrated along the $z$-direction (symmetry axis of the
particle), the term proportional to $3z^2$ dominates, $Q_0$ is
positive, and the particle is prolate shaped. If the charge
density is concentrated in the equatorial plane perpendicular to
$z$ axis, the term proportional to $r^2$ prevails, $Q_0$ is
negative and the particle is oblate in shape. The angular momentum
selection rule however, does not allow the spin $\frac{1}{2}^+$
baryons to have any spectroscopic quadrupole moment. Therefore,
the $\gamma^* N_{\frac{1}{2}^+} \to \Delta^+_{\frac{3}{2}^+}$
transition is studied to understand the quadrupole amplitudes in
the nucleon as well as $\Delta^+$.

The spin and parity conservation in the $\Delta^+ p$ transition
require three contributing amplitudes, magnetic dipole $M1$, the
electric quadrupole moment $E2$, and the charge quadrupole moment
$C2$ photon absorption amplitudes. The information on the {\it
intrinsic} quadrupole moments can be obtained from the
measurements of $E2$ and $C2$ amplitudes \cite{tia03,ber03}. If
the charge distribution of the initial and final three-quark
states were spherically symmetric, the $E2$ and $C2$ amplitudes of
the multipole expansion would be zero \cite{becchi}. However, the
recent experiments at JLAB, SELEX Collaboration reveal that
although these quadrupole amplitudes are small compared to the
dominant magnetic dipole transition $M1$, they are clearly non
zero \cite{pdg}.  More recently, the quadrupole transition moment
($Q_{\Delta^+ N}$) measured by LEGS and Mainz Collaborations is
$-0.108 \pm 0.009 \pm 0.034 ~{\rm fm}^2$ \cite{blanquad} and
$-0.0846 \pm 0.0033 ~{\rm fm}^2$ \cite{tia03}, respectively. These
measurements lead to the conclusion that the nucleon and the
$\Delta^+$ are intrinsically deformed.

\section{General Parameterization method}

In order to predict the sign as well as magnitude of deformation
in the octet and decuplet baryons, we have used the general
parameterization (GP) method \cite{morp89}. The charge quadrupole
operator composed of a two- and three-body operator terms in
spin-flavor space is given as \bea {Q} & = & {\mathrm B} \sum_{i
\ne j}^3 e_i \left( 3 \sigma_{i \, z} \sigma_{ j \,z} -
\bf{\sigma}_i \cdot \bf{\sigma}_j \right) + {\mathrm C} \!\!
\sum_{i \ne j \ne k }^3 e_i \left( 3 \sigma_{j \, z} \sigma_{ k \,
z} - \bf{\sigma}_j \cdot \bf{\sigma}_k \right) \,, \label{quad}
\eea where the coefficients ${\mathrm B}$ and ${\mathrm C} $ are
the constants to be determined from the experimental observations
on charge radii and quadrupole moments. The quadrupole moments
${Q}$ for the octet and decuplet baryons can be calculated from
Eq. (\ref{quad}) by evaluating matrix elements of the operator
corresponding to the three-quark spin-flavor wave functions  ($
Q_{B} = \left \langle {B} \vert {Q}{\vert B} \right \rangle)$. It
is straightforward to verify that \bea \sum_{i \neq j} e_i (
\xbf{\sigma_i} \cdot \xbf{\sigma_j}) &=& 2 {\xbf J} \cdot \sum_{i}
e_i \xbf{\sigma_i} - 3 \sum_{i} e_i \,, \label{eisisj} \eea \bea
\sum_{i \neq j \neq k} e_i(\xbf{\sigma_j} \cdot \xbf{\sigma_k})
&=& \pm 3 \sum_i e_i - \sum_{i \neq j} e_i( \xbf{\sigma_i} \cdot
\xbf{\sigma_j}) \,.\label{eisjsk} \eea In Eq. (\ref{eisjsk}), $+$
sign holds for $J=\frac{3}{2}$ and $-$ sign for $J = \frac{1}{2}$
states. Using the expectation value of operator $2 J \cdot
\sum_{i} e_i \xbf{\sigma_i}$ between the baryon wavefunctions $|B
\rangle$ in the initial and final states \cite{yaoubook}, the
operators in Eqs. (\ref{eisisj}) and (\ref{eisjsk}) become \bea
\ba{c|c|c} {\rm Operator} & \sum_{{i \neq j}} e_i(\xbf{\sigma_i}
\cdot \xbf{\sigma_j}) & \sum_{i\neq j \neq k} e_i(\xbf{\sigma_j}
\cdot \xbf{\sigma_k}) \\ \hline J = \frac{1}{2} & 3 \sum_{i} e_i
\xbf{\sigma_{iz}} - 3 \sum_{i} e_i & - 3 \sum_{i} e_i
\xbf{\sigma_{iz}}\\ J = \frac{3}{2} & 5 \sum_{i} e_i
\xbf{\sigma_{iz}} - 3 \sum_{i} e_i  & 6 \sum_i e_i - 5 \sum_i e_i
\xbf{\sigma_{iz}}  \\  \ea \label{tensor}\eea

The expression for the quadrupole moment of the octet and decuplet
baryons in Eq. (\ref{quad}) can be expressed as \bea {{Q}}_{1/2}
&=& 3{\mathrm B}\sum_{i \neq j} e_i \bf{\sigma_{iz}}
\bf{\sigma_{jz}} + 3 {\mathrm C}\sum_{i \ne j \ne k } e_i
\bf{\sigma_{jz}} \bf{\sigma_{kz}} +(- 3{\mathrm B} +3 {\mathrm C}
) \sum_{i} e_i \bf{\sigma_{iz}} +3 {\mathrm B}\sum_{i} e_i \,,
\label{q1/2} \\ {{ Q}}_{3/2} &=& 3 {\mathrm B} \sum_{i \neq j} e_i
\bf{\sigma_{iz}} \bf{\sigma_{jz}} + 3{\mathrm C} \sum_{i \ne j \ne
k } e_i \bf{\sigma_{jz}} \bf{\sigma_{kz}} + (-5 {\mathrm B} + 5
{\mathrm C})\sum_{i} e_i \bf{\sigma_{iz}} + (3 {\mathrm B} -6
{\mathrm C}) \sum_{i} e_i \,,\label{q3/2} \eea where $i = (u,d,s)$
for any of the three quarks. Therefore, the calculation of
quadrupole moment reduces to the calculation of the flavor
structure $(\sum_i e_{i})$, spin structure $(\sum_i e_i
\sigma_{iz})$ and the tensor terms $(\sum_i e_i
\sigma_{iz}\sigma_{jz}$ and $\sum_i e_i \sigma_{jz} \sigma_{kz})$
for a given baryon.

The spin and flavor structure of a given baryon can be calculated
using the ${\rm SU}(6)$ spin-flavor symmetry of the wave
functions. The expectation value of the appropriate operators are
expressed as $\widehat e_{i} \equiv \langle B| \sum_i e_{i}|B
\rangle \,,$ and $ \widehat {e_{i} \sigma_{iz}} \equiv \langle B|
\sum_i {e_i} {\sigma_{iz}} |B \rangle $, where $|B \rangle$ is the
baryon wave function and $e_{i}$ and $\sigma_{i}$ are the charge
and spin operators defined as \bea \sum_i e_i &=& \sum_{q=u,d,s}
n^B_{q}q + \sum_{{\bar q} = {\bar u}, {\bar d}, {\bar s}}
n^B_{{\bar q}} {\bar q} = n^B_{u}u + n^B_{d}d + n^B_{s}s +
n^B_{\bar u}{\bar u} + n^B_{ \bar d}{\bar d} + n^B_{\bar s}{\bar
s} \label{numei} \,, \\ \sum_i {e_i} {\sigma_{iz}} &=&
\sum_{q=u,d,s} (n^B_{q_{+}}q_{+} + n^B_{q_{-}}q_{-}) =
n^B_{u_{+}}u_{+} + n^B_{u_{-}}u_{-} + n^B_{d_{+}}d_{+} +
n^B_{d_{-}}d_{-} + n^B_{s_{+}}s_{+} + n^B_{s_{-}}s_{-}
\label{numeisi} \,. \eea Here $n_q^B$$(n_{\bar q}^B)$ is the
number of quarks with charge $q$(${\bar q}$), and
$n^B_{q_{+}}$($n^B_{q_{-}}$) is the number of $q$ quarks with spin
$+$($-$). The tensor terms can be simplified and further reduced
to the calculations of spin and flavor structure as presented in
Eq. (\ref{tensor}).

For the case of octet baryons, Eq. (\ref{q1/2}) can be solved for
the spin-flavor symmetric SU(6) octet baryon wave function by
using the operators  defined in Eqs. (\ref{numei}) and
(\ref{numeisi}). For ready reference, the quadrupole moment of
proton, neutron and $\Sigma^+$ are expressed as \bea {{Q}}_p &=&
3{\mathrm B} \left(- 2u_+ - d_+ + 2 u + d \right) + {\mathrm C}
\left(-4u + d + 4u_+ - d_+ \right) \,, \label{qp}
\\ {{Q}}_n &=& 3 {\mathrm B} \left( -u_+ - 2d_+ + u + 2d \right) +
{\mathrm C} \left( u - 4 d - u_+ + 4d_+ \right) \,, \label{qn} \\
{{Q}}_{\Sigma^+} &=& 3 {\mathrm B} \left( -2 u_+ - s_+ + 2u + s
\right) + {\mathrm C} \left( -4u + s + 4u_+ - s_+ \right) \,.
\label{qs} \eea Similarly, the quadrupole moment of the decuplet
baryon $\Delta^+$ can be calculated from Eq. ({\ref{q3/2}}) by
solving the matrix elements corresponding to the decuplet baryon
wave function and by using the operators defined in Eqs.
({\ref{numei}) and ({\ref{numeisi}). We have \bea {{
Q}}_{{\Delta^+}} &=& {\mathrm B}\left( 2u_+ + d_+ + 3(2 u +
d)\right) + {\mathrm C}\left( -3(2u + d) + 5(2 u_+ + d_+ )\right)
\,. \label{qd} \eea

In the case of naive quark model (NQM), the calculations lead to
$u= \frac{2}{3}$, $d = -\frac{1}{3}$, $s = -\frac{1}{3}$, $u_+ =
\frac{2}{3}$, $d_+ =- \frac{1}{3}$, and $s_+ = -\frac{1}{3}$. The
quadrupole moments of $p$, $n$, $\Sigma^+$, and $\Delta^+$ are now
given as \be {{Q}}_p =0\,, ~~~ {{Q}}_n =0 \,, ~~~{{Q}}_{\Sigma^+}
=0 \,, ~~~ {{Q}}_{{\Delta^+}} = 4{\mathrm B} + 2{\mathrm C}\,.\ee
These results suggest that the octet baryons are spherically
symmetric and there is no deformation. However, the experimental
results suggest that the nucleon is intrinsically deformed.
Therefore, there is a need to go beyond this model to include the
effects of ``quark sea''.

\section{Chiral constituent quark model}
The basic process in the $\chi$CQM is the emission of a GB by a
constituent quark which further splits into a $q \bar q$ pair as $
q_{\pm} \rightarrow {\rm GB}^{0} + q^{'}_{\mp} \rightarrow (q \bar
q^{'}) +q_{\mp}^{'}\,, \label{basic}$ where $q \bar q^{'} +q^{'}$
constitute the ``quark sea'' \cite{cheng,johan,song}. The
effective Lagrangian describing interaction between quarks and a
nonet of GBs is ${\cal L} = g_8 \bar q \Phi' q \,,$ where $g_8$ is
the coupling constant for the GBs. The GB field $\Phi'$ is
expressed in terms of the quark contents as \bea {\Phi'} = \left(
\ba{ccc} \phi_{uu} u \bar u+ \phi_{ud} d \bar d +\phi_{us} s \bar
s& \varphi_{ud} u \bar d & \varphi_{us} u \bar s
\\ \varphi_{du} d \bar u & \phi_{du}u \bar u+ \phi_{dd} d \bar d
+\phi_{ds} s \bar s & \varphi_{ds} d \bar s
\\ \varphi_{su} s \bar u & \phi_{sd} s \bar d & \phi_{su} u \bar u
+ \phi_{sd} d \bar d +\phi_{ss} s \bar s \\ \ea \right)\,, \eea
where \bea \phi_{uu} &=& \phi_{dd}= \frac{1}{2} +\frac{\beta}{6} +
\frac{\zeta}{3}\,,~~~~~ \phi_{ss} =\frac{2\beta}{3} +
\frac{\zeta}{3}\,,~~~~~ \phi_{us} = \phi_{ds}=
\phi_{su}=\phi_{sd}= -\frac{\beta}{3} + \frac{\zeta}{3}\,,
\nonumber \\ \phi_{du} &=& \phi_{ud}= -\frac{1}{2}
+\frac{\beta}{6} + \frac{\zeta}{3}\,,~~~~~ \varphi_{ud} =
\varphi_{du} = 1\,, ~~~~\varphi_{us} = \varphi_{ds}=\varphi_{su} =
\varphi_{sd}= \alpha \,. \label{phi1} \eea SU(3) symmetry breaking
is introduced by considering $M_s > M_{u,d}$ as well as by
considering the masses of GBs to be nondegenerate $(M_{K, \eta} >
M_{\pi}$ and $M_{\eta^{'}} > M_{K, \eta})$
{\cite{cheng,johan,song,hd}. The parameter $a(=|g_8|^2$) denotes
the transition probability of chiral fluctuation of the splitting
$u(d) \rightarrow d(u) + \pi^{+(-)}$, whereas $\alpha^2 a$,
$\beta^2 a$ and $\zeta^2 a$ respectively, denote the probabilities
of transitions of~ $u(d) \rightarrow s  + K^{-(o)}$, $u(d,s)
\rightarrow u(d,s) + \eta$, and $u(d,s) \rightarrow u(d,s) +
\eta^{'}$.

Since the quadrupole moment operators for the spin $\frac{1}{2}$
and spin $\frac{3}{2}$ baryons involve the knowledge of spin and
flavor structure of baryons, it is important to mention here that
redistribution of flavor and spin take place among the ``quark
sea''. The modified flavor and spin structure of the baryon in
$\chi$CQM due to the chiral symmetry breaking can be calculated by
substituting for every constituent quark \bea q \to P_q q +
|\psi(q)|^2  ~~~~~~~~ {\rm and} ~~~~~~~~q_{\pm} \to P_q q_{\pm}+
|\psi(q_{\pm})|^2 \,, \label{spin} \eea where $P_q = 1 - \sum P_q$
is the transition probability of no emission of GB from any of the
$q$ quark, $|\psi(q)|^2$ is the transition probability of the $q$
quark, and $|\psi(q_{\pm})|^2$ is the probability of transforming
a $q_{\pm}$ quark \cite{hd}.

\section{Results and Discussion}

In order to calculate the quadrupole moments of the octet and
decuplet baryons in the $\chi$CQM, we substitute Eq. (\ref{spin})
in Eqs. (\ref{q1/2}) and (\ref{q3/2}) . For example, on
substituting Eq. (\ref{spin}) in Eqs. (\ref{qp}), (\ref{qn}) and
(\ref{qs}), the quadrupole moments of $p$, $n$, and $\Sigma^+$ in
$\chi$CQM can now be expressed as \bea {{Q}}_p &=& {\mathrm B} a
\left(6 + \beta^2 + 2 \zeta^2 \right) - {\mathrm C} a \left(4 + 2
\alpha^2 + \beta^2 + 2 \zeta^2 \right) \,, \label{qp1}
\\ {{Q}}_n &=& {\mathrm B} a (3 - 3 \alpha^2) + {\mathrm C}
\frac{a}{3}(3 + 9 \alpha^2 + 2 \beta^2 + 4 \zeta^2) \,,
\label{qn1} \\ {Q}_{\Sigma^+}  &=& {\mathrm B}a (6 + \alpha^2 + 2
\zeta^2) -{\mathrm C} \frac{a}{3} (5 \alpha^2 + 4 \beta^2 + 12 + 6
\zeta^2) \,.\label{qs1} \eea From the above equations, we can
directly estimate the effects of SU(3) symmetry breaking and pion
cloud in the quadrupole moments of the octet baryons. It is clear
from the equations that, for the non zero value of the GP
parameters ($\mathrm B$ and $\mathrm C$), there is a significant
contribution of the $\chi$CQM parameters ($a, \alpha, \beta,
\zeta$) to the quadrupole moments. Similarly, the quadrupole
moment of $\Delta^+$ and $\Xi^{*0}$ in $\chi$CQM, after
substituting the contribution coming from the ``quark sea'', can
be expressed as \bea {{ Q}}_{{\Delta^+}} &=& {\mathrm B}\left(4 -
\frac{1}{3} a \left(6 + \beta^2 + 2 \zeta^2 \right) \right) +
{\mathrm C} \left(2 - \frac{5}{3} a \left(6 + \beta^2 + 2\zeta^2
\right) \right) \,, \label{qd1} \\ {Q}_{\Xi^{*0}} &=& {\mathrm B}
\frac{a}{3} (-3 + \alpha^2 + 2 \beta^2) + {\mathrm C} \frac{5a}{3}
(-3 + \alpha^2 + 2 \beta^2)\,. \eea

Before giving the numerical results, we would like to discuss the
input parameters involved in the calculations of baryon quadrupole
moments in $\chi$CQM. The calculations involve the two set of
parameters, the ones corresponding to the GP method (${\mathrm B}$
and ${\mathrm C}$, where ${\mathrm B}> {\mathrm C}$), and other
corresponding to the $\chi$CQM symmetry breaking ($a$, $\alpha$,
$\beta$, and $\zeta$). In order to fix the values of GP
parameters, we have performed a fit to the available experimental
values of the baryon charge radii leading to $ {\mathrm B} =
-0.0525 \,, ~~ {\mathrm C} = -0.0158 \,,$ as the best fit. For the
$\chi$CQM parameters, we have used the same set of parameters as
discussed in our earlier publication \cite{nsweak}. The values
used are $ a = 0.12\,, ~~~ \alpha = 0.45\,, ~~~ \beta = 0.45\,,
~~~ \zeta = -0.15$.

Using the above discussed set of parameters, we have calculated
the quadrupole moments of octet and decuplet baryons in $\chi$CQM
and results have been presented in Tables \ref{quad1/2} and
\ref{quad3/2}. In order to understand the role of chiral symmetry
breaking and SU(3) symmetry breaking, we have also presented the
results of NQM in the tables.  Most of the models in literature
are unable to estimate the {\it intrinsic} quadrupole moment of
the baryons. In the case of octet baryons there are indirect
evidences of small deformation in the nucleon. This deformation
can easily be observed in our results. For the case of decuplet
baryons also, $\chi$CQM is able to give a fairly good prediction
of the sign as well as magnitude of the quadrupole moments.

In order to understand the effect of three-quark contributing term
and to make our calculations more responsive, we have also
presented the results by neglecting this contribution which can be
obtained by fixing the coefficient $\mathrm C =0$. From Table
\ref{quad1/2}, it is clear that the results are affected to a very
small extent by the inclusion of three-quark term. The three-quark
terms do not seem to play any significant role in the case of
octet baryons. However, in the case of decuplet baryons, the
results in Table \ref{quad3/2} reveal that the inclusion of
three-quark term increases the quadrupole moments thus making them
significant in this case. It is interesting to observe that the
effect of three-quark contribution is even more in the case of
neutral baryons. For example, in the case of charged baryons, the
predictions are increased about 10\% whereas for the case of
neutral particles the variation is more then 50\% making the
effect of three-quark term significant.

Since the electric quadrupole moment of the octet baryons as well
as that of octet-decuplet transitions are amenable to measurement,
any experimental information would have important implications for
the basic tenets of $\chi$CQM and the effects of SU(3) symmetry
breaking.

\begin{table}
\begin{center}
\begin{tabular}{|c|c|c|}
\hline
Baryon&\multicolumn{2}{c|}{$\chi$CQM} \\ \hline
&  ${\mathrm C}$ = 0 & ${\mathrm C}$ = $-$0.016 \\ \hline ${
Q}_{p}$ & $-$0.041& $-$0.032  \\ ${ Q}_{n}$ &$-$0.016 & $-$0.019
\\ ${ Q}_{\Sigma^{+}}$ &$-$0.041 & $-$0.032 \\ ${ Q}_{\Sigma^{-}}$
&0.010 & 0.009 \\ ${ Q}_{\Sigma^{0}}$ &$-$0.016& $-$0.012 \\ ${
Q}_{\Xi^{0}}$ & $-$0.016 & $-$ 0.019\\ ${ Q}_{\Xi^-}$ & 0.010&
0.009 \\
\hline
\end{tabular}
\end{center} \caption{Quadrupole moments of spin $\frac{1}{2}^+$
baryons in $\chi$CQM. The deformation is zero for all the cases in
NQM. }\label{quad1/2}
\end{table}

\begin{table}
\begin{center}
\begin{tabular}{|c|c|c|c|}
\hline Baryon &  NQM& \multicolumn{2}{c|}{$\chi$CQM}  \\ \hline &
& ${\mathrm C}$ = 0 &${\mathrm C}$ = $-$0.016 \\ \hline ${
Q}_{\Delta^{++}}$  & $-$0.484 & $-$0.398 &$-$0.428 \\
${Q}_{\Delta^{+}}$ & $-$0.242 &$-$0.196 &$-$0.208 \\ ${
Q}_{\Delta^{0}}$ & 0.0 & 0.005 & 0.013 \\ ${ Q}_{\Delta^{-}}$
&0.242 & 0.207 & 0.208 \\ ${Q}_{\Sigma^{*+}}$ &$-$0.242 &$-$0.196
&$-$0.208 \\ ${ Q}_{\Sigma^{*-}}$ &0.242 & 0.207 & 0.234\\ ${
Q}_{\Sigma^{*0}}$ &0.0 & 0.005 &0.013 \\
 ${Q}_{\Xi^{*0}}$ & 0.0& 0.005 & 0.013 \\
${ Q}_{\Xi^{*-}}$ &0.242 & 0.207 & 0.234\\ ${ Q}_{\Omega^-}$ &
0.315 & 0.296 &0.296 \\
  \hline
\end{tabular}
\end{center} \caption{Quadrupole moments of spin $\frac{3}{2}^+$ baryons in $\chi$CQM.}
\label{quad3/2}
\end{table}

\end{document}